# Classification of Functioning, Disability, and Health for Children and Youth: ICF-CY Self Care (SCADI Dataset) Using Predictive Analytics

Abstract ID: 570180

Avishek Choudhury
School of Systems and Enterprises
Stevens Institute of Technology
Hoboken, NJ

Dr. Christopher M. Greene
Binghamton University
Binghamton, NY

## Abstract

The International Classification of Functioning, Disability, and Health for Children and Youth (ICF-CY) is a scaffold for designating and systematizing data on functioning and disability. It offers a standard semantic and a theoretical foundation for the demarcation and extent of wellbeing and infirmity. The multidimensional layout of ICF-CY comprehends a plethora of information with about 1,400 categories, making it difficult to analyze. Our research proposes a predictive model that classifies self-care problems on Self-Care Activities Dataset based on the ICF- CY. The data used in this study is comprised of 206 attributes of 70 children with motor and physical disability. Our study implements, compares and analyzes Random Forest, Support vector machine, Naïve Bayes, Hoeffding tree, and Lazy locally weighted learning using two-tailed T-test at 95% confidence interval. Boruta algorithm involved in the study minimizes the data dimensionality to advocate the minimal-optimal set of predictors. Random forest gave the best classification accuracy of 84.75%; root mean squared error of 0.18 and receiver operating characteristic of 0.99. Predictive analytics can simplify the usage of ICF-CY by automating the classification process of disability, functioning, and health.

## Introduction

The International Classification of Functioning, Disability, and Health for Children and Youth (ICF-CY) is a derivative from the International Classification of Functioning, Disability, and Health (ICF) and is intended to record the characteristics of a developing child and the influence of its surrounding environment. ICF-CY is not just a classification, but it introduces a biopsychosocial model to document body functions and structures, activities and participation of children and youth, and their environments across developmental stages. The ICF is a scaffold for designating and systematizing data on functioning and disability. It offers a standard semantic and a theoretical foundation for the demarcation and extent of wellbeing and infirmity. In 2001, the World Health Assembly sanctioned the ICF. An acquaintance taxonomy for children and youth (ICF-CY) was broadcasted in 2007. The ICF-CY amalgamates the significant disability facsimiles. It acknowledges the role of environmental influences in the foundation of disability and the consequences of allied health conditions. The ICF-CY encompasses multidimensional concepts related to functioning and disability. It relates to the **(a)** physical functions and edifices of an individual, **(b)** the activities performed by an individual and their constraints, **(c)** restrictions experienced by individuals in diverse fields, and **(d)** the environmental influences [1]. Disability is the consequence of physical or psychological ailments that hinder human performance. Physical and motor frailties are conditions that impede people's actions [2]. Like all other health concerns, identification and classification of an individual's frailty is an exhaustive course that compels specialists, psychotherapists, and doctors. The shortage of expert therapists amplifies wait time and work stress, therefore making the classification challenging. An individual's functioning is a vibrant interface between her or his health, environmental elements, and private characteristics. ICF-CY is a biopsychosocial archetypal of disability that incorporates the communal and medical models [1]. **Figure 1** [1] below shows an example of the interaction between various components of ICF-CY.

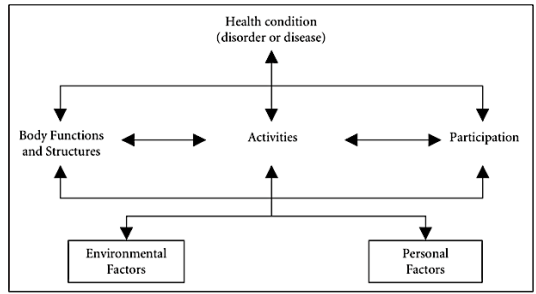

Figure 1: Interaction among components of ICF-CY

ICF-CY is designed framework which is globally used for disability classification. ICF-CY is primarily categorized into (a) body function, (b) body structure, (c) activities and participation, and (d) environmental factors. Each of these is again subclassified as shown in Table 1 below [1].

Table 1: ICF-CY Primary categories and subclassification

| Body Function | Body Structure | Activities and Participation | Environmental Factors |
|---|---|---|---|
| Mental functions | Structure of the nervous system | Learning and applying knowledge | Products and technology |
| Sensory roles and pain | The eye, ear and related structures | | |
| | | General tasks and demands | Natural environment and human-made changes to environment |
| Voice and speech functions | Structures involved in voice and speech | | |
| Utilities of the cardiovascular, hematological, immunological and respiratory systems | Structure of the cardiovascular, immunological and respiratory systems | Communication and mobility | |
| | | Self-care | |
| Functions of the digestive, metabolic, and endocrine systems | Structures related to the digestive, metabolic, and endocrine systems | Domestic life | Support and relationships |
| | | Interpersonal interactions and relationships | Attitudes, services, systems, and policies |
| Genitourinary and reproductive functions, neuro-musculoskeletal and movement-related functions | Structures related to genitourinary and reproductive systems | | |
| | Structures related to movement | Major life areas | |
| Functions of the skin and related structures | Skin and related structures | Community, social and civic life | |

For a more in-depth analysis of all ICF-CY's four broad categories, each ICF-CY sub-classifications are demarcated using ICF Qualifier scales as shown in Table 2 below [1].

Table 2: ICF Qualifier Scales

| Generic Qualifiers | Qualifiers for Environmental Factors | |
|---|---|---|
| 0 No problem | 0 No barrier | +0 No facilitator |
| 1 Mild problem | 1 Mild barrier | +1 Mild facilitator |
| 2 Moderate problem | 2 Moderate barriers | +2 Moderate facilitator |
| 3 Severe problem | 3 Severe barriers | +3 Severe facilitator |

| 4 Complete problem | 4 Complete barriers | +4 Complete facilitator |
|---|---|---|
| 8 Not specified | 8 Not specified | +8 Facilitator, not specified |
| 9 Not applicable | 9 Not applicable | +9 Not applicable |

The ICF and ICF-CY are acknowledged as a component of the World Health Organization Family of International Classifications (WHO-FIC), and a harmonizing member of the International Statistical Classification of Diseases and Related Health Problems (ICD) [1]. Since health conditions, disabilities, and environmental factors are all interdependent and interrelated, ICF, if used with ICD will generate a meaningful classification of functionality and health conditions [1]. Since 2007, ICF-CY is not updated by WHO; however, ICF-CY being a derived classification of ICF, includes additional detailed information on the application of the ICF when supporting the pertinent aspects of functioning and health in children and youth [3]. In FDRG meeting held at Madrid in June 2010, the merger of ICF-CY and ICF was recommended approved in October 2010 at the WHO-FIC Toronto Meeting. The proposed merger of the ICF and ICF-CY establishment will have significant implications on the classification and maintenance and will enhance the ICF coverage [3]. The ongoing merging process mandates additional items to be in ICF foundation layer to match ICF-CY [3]. It is believed that linearized expression of ICF-CY will allow simpler maintenance, translation, and printing, as well as compatibility and inclusion, in electronic health applications [3]; however, the problems associated with more extensive and complex data structure is not addressed.

## Problem Statement

For accurate and meaningful classification of disability, health condition, and their relationship, acknowledging all constituents of the ICF-CY are advised. Such classification structures have been pronounced as the foundation of statistical evidence. The ICF-CY alone comprises approximately 1,400 categories. Its expansive and multidimensional structure makes it difficult and exhaustive to perform any classification and statistical analysis. To develop an accurate and faster decision support system, this study employed machine learning algorithms on SCADI dataset. This approach streamlines and automates the health and disability classification process and consecutively provides ICF-CY users (doctors, lawyers, policymakers, researchers and others) an easy to use classification tool.

## Methodology

The data used in this study was donated on 14[th] April 2018 by Dr. Fatemi Bushehri [4], Department of Software Engineering, Yazd Branch, Islamic Azad University, Yazd, Iran [4]. It consists of 206 attributes including *one* response variable ("class") and *70* instances. The data used in this study contains only the 'self-care' (as highlighted in table 1) classification under 'Activities and Participation' category of ICF-CY. The "class" field refers to the presence of the self-care problems of children with physical and motor disabilities. Occupational therapists determine the classes. The predictors consist of gender (*1 = male; 0 = female*), age in years, self-care activities based on ICF-CY (*1 = the child has this feature; 0 = otherwise*), and a response variable class ( *class1 = Caring for body parts problem; class2 = Toileting problem; class3 = Dressing problem; class4 = Washing oneself and Caring for body parts and Dressing problem; class5 = Washing oneself, Caring for body parts, Toileting, and Dressing problem; class6 = Eating, Drinking, Washing oneself, Caring for body parts, toileting, Dressing, Looking after one's health and Looking after one's safety problem; class7 = No Problem* )

To develop the best-fit classification model, this study conducts a comparative analysis of prediction models and implements a feature selection method for further optimization. We imported the raw data into Weka 3.8.2 for all comparative analysis. Feature selection using Boruta algorithm was steered in RStudio 1.1.456. Feature selection is an essential step in machine learning methods. Datasets are often designated with too many variables for efficient model structure [5]. Commonly, most of these variables are extraneous to the classification, and perceptibly their relevance is unknown in advance [5]. Several difficulties are dealing with large feature sets. One is essentially technical — dealing with large feature sets impedes computational speed, consumes too many resources and is merely bothersome. Another is even more important — many machine learning algorithms reveal a diminution of accuracy when the number of variables is considerably higher than optimal [6]. Therefore, selection of minimal feature set that can yield the best possible classification outcome is needed for practical reasons [5]. This problem is also known as minimal-optimal problem; it has been intensively analyzed and there are several algorithms which are established to reduce the feature set to a manageable and optimal size [5].

Nevertheless, this genuine goal sleuths another problem — the identification of all attributes which are in certain circumstances germane for classification, the so-called "all-relevant problem" [5]. Finding all relevant attributes, instead of the non-redundant ones, may be beneficial. For example, when dealing with classification of SCADI dataset, identification of all predictors which are related to the outcome ("class1" through "class7") is necessary for complete understanding of the process, whereas a minimal-optimal set of predictors (variables) might be more useful as classification markers. An honest discussion demarcating the importance of finding relevant attributes is given by Nilsson et al. in 2007. The dilapidation of the classification accuracy, upon elimination of the features from a feature set, is ample to declare the feature essential; however, lack of this effect is not satisfactory to proclaim it as unimportant.

One therefore needs another criterion for declaring variables essential or unimportant. In a wrapper method, the classifier is used as a black box returning the feature ranking; therefore, one can use any classifier which can provide the ranking of features [5]. Boruta algorithm is a wrapper built around the random forest classification algorithm [5] implemented in the R package random Forest. Boruta algorithm uses Z score as the measure of importance since it considers the fluctuations of the mean accuracy loss among trees in the forest [5]. Since we cannot use Z score unswervingly to gauge importance, an external reference is needed to decide whether the importance of any given attribute is significant. To determine the importance of each attribute, Boruta algorithm creates an analogous 'shadow' attribute, whose values are obtained by shuffling values of the original attribute across objects [5]. Then, a classification is performed using all the attributes of the extended system to calculate the importance of all attributes. The importance of a shadow attribute can be nonzero purely due to random fluctuations [5]. Thus, the set of the importance of shadow attributes is used as a reference for determining essential attributes.

This study implements k-fold cross-validation to minimize any bias and variance in the dataset. Cross-validation is a resampling technique used to gauge machine learning models on a limited dataset. In this method, the original data sample is randomly partitioned into *k* equal subsamples. Of the *k* subsamples, one subsample is retained as the validation data for evaluating the model, and the remaining *k-1* subsamples are used as training data. The cross-validation process is then reiterated *k* times. The *k* results obtained from the *k*-folds are then averaged to produce a single estimation. In this study, we considered the value of *k* to be 10 becoming 10-fold cross-validation. The following machine learning algorithms were implemented, compared and tested using two-tailed corrected paired T-test at 0.05 significance.

**Random Forest –** Also known as random decision forest, it is an ensemble method used to construct predictive models for both classification and regression problems. Ensemble methods employ multifarious learning models to gain enhanced predictive outcome — regarding random forest, the model creates a forest of random uncorrelated decision trees to attain the best possible riposte.

**Support Vector Machine –** A Support Vector Machine (SVM) is a discriminative classifier formally expressed by a parting hyperplane. In other words, under supervised learning, the algorithm yields an optimal hyperplane which compartmentalizes new instances. In two-dimensional space, this hyperplane is a line separating a plane in two segments were each class lay on either side.

**Naïve Bayes –** Naïve Bayes classifier is a group of probabilistic classifiers established on Bayes' theorem with the active (naïve) assumption of independence between the attributes. It is favorably scalable, obliging several parameters linear in the number of predictors in a learning problem. This classifier can be trained proficiently in a supervised learning situation. In numerous applications, parameter estimation for naive Bayes models involves maximum likelihood; in other words, the Naïve Bayes model does not mandate the use of Bayesian probability or any Bayesian methods.

**Hoeffding tree -** Also known as Very Fast Decision Tree (VFDT), it is a tree algorithm for data stream classification. The Hoeffding tree is an incremental decision tree learner for a large dataset, that assumes that the data distribution is constant over time. It grows a decision tree based on the theoretical guarantees of the Hoeffding bound. In other words, VFDT employs Hoeffding bound to decide the minimum number of arriving instances to achieve a certain level of confidence in splitting the node. The confidence level determines the proximity of the statistics between the attribute chosen by VFDT and the attribute chosen by decision tree for batch learning.

**Lazy locally weighted learning** - It is a class of non-parametric function approximation techniques, where a prediction is made by using an approximated local model around the current point of interest.

While evaluating supervised machine learning models, it is important to measure each classification of the model in *accuracy; root mean squared error* (RMSE), and *receiver operating characteristic* (ROC). Classification accuracy is the metric for evaluating classification models. It is the fraction of predictions or classification that a model performs correctly. Classification accuracy can be calculated by the given equation (eq.1)

$$Accuracy = \frac{Number\ of\ correct\ prediction}{Total\ number\ of\ prediction} = \frac{TP + TN}{TP + TN + FP + FN} \tag{1}$$

Where *TP* = positive; *TN* = True negative; *FP* = False positive; *FN* = False negative.

The ROC curve is the graphical representation of the true positive rate (TPR) against the false positive rate (FPR) at different threshold settings. In the machine learning domain, a TPR is also known as sensitivity, recall or "probability of detection." Similarly, an FPR is known as the fall-out or "probability of false alarm" and can be calculated as (eq. 2). The ROC curve is thus the sensitivity as a function of fall-out.

$$FPR = (1 - specificity) \tag{2}$$

The root-mean-square error (RMSE) is a measure of performance of a model. It does this by computing the difference between predicted and the actual values as given below (eq. 3).

$$RMSE = \sqrt{\sum_{i=1}^{N} \frac{(x_i - y_i)^2}{N}} \tag{3}$$

Where $(x_i - y_i)$ is the difference between predicted and actual value and N is the sample size.

## Results
### Performance evaluation 1
After implementing the mentioned algorithms, the following results were observed as shown in *Table 3*- below:

**Table 3:** Model performance before feature selection

|  | Random Forest | Support Vector Machine | Naïve Bayes | Lazy LWL | Hoeffding tree |
|---|---|---|---|---|---|
| Accuracy | **83.28 %** | 78.86 % | 82.57 % | 78.71 % | 82.57 % |
| RMSE | **0.18** | 0.22 | 0.19 | 0.22 | 0.19 |
| ROC | **0.84** | 0.80 | 0.61 | 0.52 | 0.67 |

Boruta algorithm identifies 55 as significant variables. This technique reduces the dataset by approximately 74% without affecting prediction accuracy.

### Performance evaluation 2
The section only considers the selected important attributes for comparative analysis and shows the comprehended result in table 4 given below. All the algorithms and performance measures remain the same.

**Table 4:** Comparative analysis using only selected important attributes

|  | Random Forest | Support Vector Machine | Naïve Bayes | Lazy LWL | Hoeffding tree |
|---|---|---|---|---|---|
| Accuracy | **84.75%** | 82.43% | 83.00% | 78.57 % | 80.71% |
| RMSE | **0.18** | 0.20 | 0.19 | 0.21 | 0.21 |
| ROC | **0.99** | 0.96 | 0.98 | 0.94 | 0.96 |

## Conclusion

The Random Forest algorithm performed better than the rest models. This study suggests implementing random forest on a similar data type. After removing unimportant attributes from the dataset, there was a significant improvement observed in the area under the ROC curve for all models. Moreover, no other performance measures were impeded. This study generates an accuracy of 84.75% which is so far the highest achieved accuracy on the SCADI dataset. ICF-CY is designed for healthcare, education, research, policymakers, and many other diverse fields but its complexity might deter people from using it to its full potential. ICF [1] has become a standard classification for the depiction and assessment of health, in which functioning can be implied as the operationalization of health, and exemplifies the outcome of the interaction between a person's health condition and contextual factors. The ICF encompasses 1,400 categories making its applicability challenging in everyday clinical practice. To address this concern and facilitate its broader accomplishment in numerous settings, the WHO and the ICF Research Branch crafted a scientifically based process for developing core sets of ICF categories for specific purposes [7].

An ICF Core Set (ICFCS) is a miscellany of essential categories from the ICF classification that are considered most pertinent to describe the functioning of a person with a specific health condition or in a specific healthcare context [7]. Machine learning has been performing well enough in diagnosing diseases and classifying patients based on their health conditions [8-12]. Future development and modifications in ICF-CY, particularly for healthcare providers, may consider incorporating feature selection techniques and association rules in the ICF-CF design.

## References

[1] World Health Organization, International Classification of Functioning, Disability and Health-Children & Youth Version, Switzerland: WHO, 2007.
[2] B. R. Lewis and R. Turner, "Physical disability and depression: clarifying racial/ethnic contrasts," J. Aging Health, vol. 22, no. 7, p. 977–1000, 2010.
[3] World Health Organization, "Implementing the merger of the ICF and ICF-CY," Friday, March 2018. [Online]. Available: https://www.who.int/classifications/icf/whoficresolution2012icfcy.pdf. [Accessed Tuesday, December 2018].
[4] Zarchi, Bushehri, Fatemi, and Dehghanizadeh, "SCADI: A standard dataset for self-care problems classification of children with physical and motor disability," International Journal of Medical Informatics, 2018.
[5] B. K. Miron and R. R. Witold, "Feature Selection with the Boruta Package," Journal of Statistical Software, vol. 36, no. 11, pp. 2-13, 2010.
[6] K. Ron and H. J. George, " Wrappers for feature subset selection," Artificial Intelligence, vol. 97, pp. 273-324, 1997.
[7] S. Melissa, S. Gerold, K. Nenad, U. Tevfik, et. al., "A guide on how to develop an International Classification of Functioning, Disability and Health Core Set," European journal of physical and rehabilitation medicine, pp. 1973-9095, 2014.
[8] Choudhury and Wesabi, "Classification of Cervical Cancer Dataset," in Proceedings of the 2018 IISE Annual Conference, Orlando, 2018.
[9] Choudhury and C. M. Greene, "Prognosticating Autism Spectrum Disorder Using Artificial Neural Network Levenberg-Marquardt Algorithm," Archives of Clinical and Biomedical Research, vol. 2, no. 6, pp. 188-197, 26 November 2018.
[10] E. Khan and A. Choudhury, "Decision Support System for Renal Transplantation," in Proceedings of the 2018 IISE Annual Conference, Orlando, 2018.
[11] Choudhury, "Identification of Cancer: Mesothelioma's Disease Using Logistic Regression and Association Rule," American Journal of Engineering and Applied Sciences, vol. 11, no. 4, pp. 1310-1319, 2018.
[12] Choudhury and C. M. Greene, "Evaluating Patient Readmission Risk: A Predictive Analytics Approach," American Journal of Engineering and Applied Sciences, vol. 11, no. 4, pp. 1320-1331, 2018.